\documentclass[%
 reprint,
 amsmath,amssymb,
 aps,
]{revtex4-1}
\usepackage{graphicx}% Include figure files
\usepackage{dcolumn}% Align table columns on decimal point
\usepackage{bm}% bold math
\usepackage{upgreek}% Include figure files

\begin{document}

%\preprint{}

\title{Compact multi-fringe interferometry with sub-picometer precision}


\author{Katharina-Sophie Isleif}
  \email{katharina-sophie.isleif@aei.mpg.de}

\author{Gerhard Heinzel}
\author{Moritz Mehmet}

\author{Oliver Gerberding}
 \email{contact@olivergerberding.com}

 \affiliation{Max Planck Institute for Gravitational Physics (Albert Einstein Institute)\\ and Leibniz Universit\"at Hannover\\Callinstr. 38, 30167 Hannover, Germany\\}
%\thanks{}


\date{\today}

\begin{abstract}
%Deep frequency modulation interferometry makes compact optical metrology systems with multi-fringe dynamic range feasible. 

Deep frequency modulation interferometry combines optical minimalism with multi-fringe readout. Precision however is key for applications such as optical gradiometers for satellite geodesy or as dimensional sensor for ground-based gravity experiments. 
We present a single-component interferometer smaller than a cubic inch. Two of these are compared to each other to demonstrate tilt and displacement measurements with a precision of less than $20\,\mathrm{nrad}/\sqrt{\mathrm{Hz}}$ and $1\,\mathrm{pm}/\sqrt{\mathrm{Hz}}$ at frequencies below $1\,\mathrm{Hz}$. 

%Deep frequency modulation is a self-homomdyning interferometer technique that provides the feasibility of compact optical metrology systems with a dynamic range over multiple fringes. We represent the design of a 25\,mm-large single-component interferometer. Two of these are used to demonstrate the ability of measuring tilt and displacement information of a test mass mirror with a precision of better than 20\,nrad/$\sqrt{\mathrm{Hz}}$ and 1\,pm/$\sqrt{\mathrm{Hz}}$, respectively, at frequencies below 1\,Hz. 
%The results prove the readiness of deep frequency modulation interferometry for applications that benefit from miniaturised multi-fringe optical sensors. Prominent candidates are for example optical gradiometers that would enhance Earth gravity field measurements, or inertial sensors that are used to measure ground motions. 
\end{abstract}
%% Abstract no longer than 600 characters
%% 3750 words, about 4 pages, and a Comment should not exceed 750 words, about 1 page

%% Displayed Math
% The word equivalent for displayed math is 16 words per row for single-column equations. Two-column equations count as 32 words per row.

%% Figures
% Estimating the word equivalent for figures can be simplified by using the aspect ratio (width / height) of the figure. The estimates would be [(150 / aspect ratio) + 20 words] for single-column figures, and {[300 / (0.5 * aspect ratio)] + 40 words} for double column figures.

\pacs{}
%\keywords{}

\maketitle



%%%%%%%%%%%%%%%%%%%%%%%%%%%%%%%%%%%%%%%%%%%%%%%%%%%%%%%%%%%%%%%%%%%%%%%%%%%%%%%%%%
\section{Introduction}
%%%%%%%%%%%%%%%%%%%%%%%%%%%%%%%%%%%%%%%%%%%%%%%%%%%%%%%%%%%%%%%%%%%%%%%%%%%%%%%%%%
Heterodyne laser interferometry, as used very successfully in the LISA Pathfinder mission \cite{armano2018beyond}, provides sub-picometer test mass displacement and sub-nanoradian tilt sensing sensitivities, at readout frequencies, between 1\,Hz and 1\,mHz, while operating over a large dynamic range (multiple fringes). Many applications and experiments would benefit from incorporating multiple sensors with this precision, such as future satellite geodesy missions with multiple test masses \cite{Drinkwater2006}, accelerometers \cite{cooper2018compact} and gravitational wave detectors \cite{LISAmissionL3,Dahl2012,abbott2017gw170814}. Further applications can also make use of this precision to measure any other derived quantity at low readout frequencies, either as core part of larger instruments or as auxiliary sensors to monitor, for example, thermal expansion or parasitic motions, without requiring sub-fringe stability of the objects under test. Currently though, the availability and usability of such interferometric sensors is limited by their optical complexity. The need for intricate design, the construction of ultra-stable optical benches with multiple components, many of which need individual alignment and bonding, and multiple fiber connections make their application very elaborate. These factors drive the instrument size, mass, construction effort and lead time and limit the current range of deployment significantly.

%The complexity of scientific instruments, including factors such as their mass, size and construction effort, determines their availability and usability. The measurement quality of base units, such as length or acceleration and gravity variations, is still in high demand. 
%While most of nowadays research is about improving the measurement performance, we focussed on simplifying the instruments and achieving more minimalistic systems, which is also a form of quality enhancement. 
%
%Laser interferometers are one of the best tools to measure the base units with high accuracy. Prime examples are the ground-based gravitational wave (GW) detectors LIGO and Virgo, detecting GWs with a strain sensitivity of $10^{-21}\sqrt{\mathrm{Hz}}$ \cite{abbott2017gw170814}; or the planned space-based detector LISA \cite{LISAmissionL3}, whose Pathfinder mission sensed displacement variations of 30\,fm/$\sqrt{\mathrm{Hz}}$ at frequencies below 1\,Hz in space \cite{armano2018beyond}. 
%The extreme complexity makes an adoption of these instruments challenging for other applications.

Hence, alternative interferometer techniques that require simpler optical set-ups with fewer optical components \cite{watchi2018contributed} have the potential to reach a wider scope. Self-homodyning interferometers with phase-shift keying methods like digitally enhanced interferometry \cite{sutton2012digitally}, deep phase modulation \cite{schwarze2014advanced,heinzel2010deep} or deep frequency modulation interferometry (DFMI) \cite{Gerberding2015,1742-6596-716-1-012008,Isleif:16} are attractive techniques that reduce the optical complexity at the expense of more sophisticated phase extraction algorithms. 
In recent years many proof-of-principle studies \cite{shaddock2007digitally,sutton2012digitally,isleif2014highspeed,Isleif:16,Kissinger2013fibersegment,kissinger2015range,de2009picometer} have revealed greatly simplified optical readout schemes that rely on such hybrid approaches. Effectively they combine continuous-wave laser sources with specific modulations to create pulsed- or comb-like fields that enable optical simplification, phase readout and sometimes even multiplexing. However, probing and achieving the low-frequency displacement noise, required for example for future gravity experiments, has remained rare for these techniques \cite{de2009picometer}.

We have focused on studying DFMI, a technique that has recently gained wide interest \cite{akbarzadeh2017low,Ni:17,Zhang:17,arablu2018polydyne}. Compared with the classic heterodyne scheme or Homodyne Quadrature Interferometry \cite{cooper2018compact}, DFMI requires less optical components which allows for more compact layouts without giving up the multi-fringe capabilities. Previous studies have verified this functionality \cite{Isleif:16} and here we present an experiment that demonstrates the actual displacement sensing performance of DFMI on the 1\,pm/$\sqrt{\mathrm{Hz}}$-level, making it an attractive alternative to heterodyne interferometry even for the most sensitive gravity experiments.

%The combination of minimalistic design and high dynamic range makes DFMI usable for applications that require a very accurate multi-degree-of-freedom readout of multiple objects. Future GOCE-like satellite geodesy missions can be equipped with optical gradiometers and would allow measurements of the Earth's gravitational field with enhanced resolution \cite{Drinkwater2006}. This would show us a deeper insight into the effects of climate change like ocean rise, drawdowns and Earth quakes. 

% 150 mm x 51.329 mm 
% ratio = 2.9223246118
% words = 300 / (0.5*2.9223246118) + 40 = 245.3160000013
 \begin{figure*}[]
\includegraphics[width=0.95\textwidth]{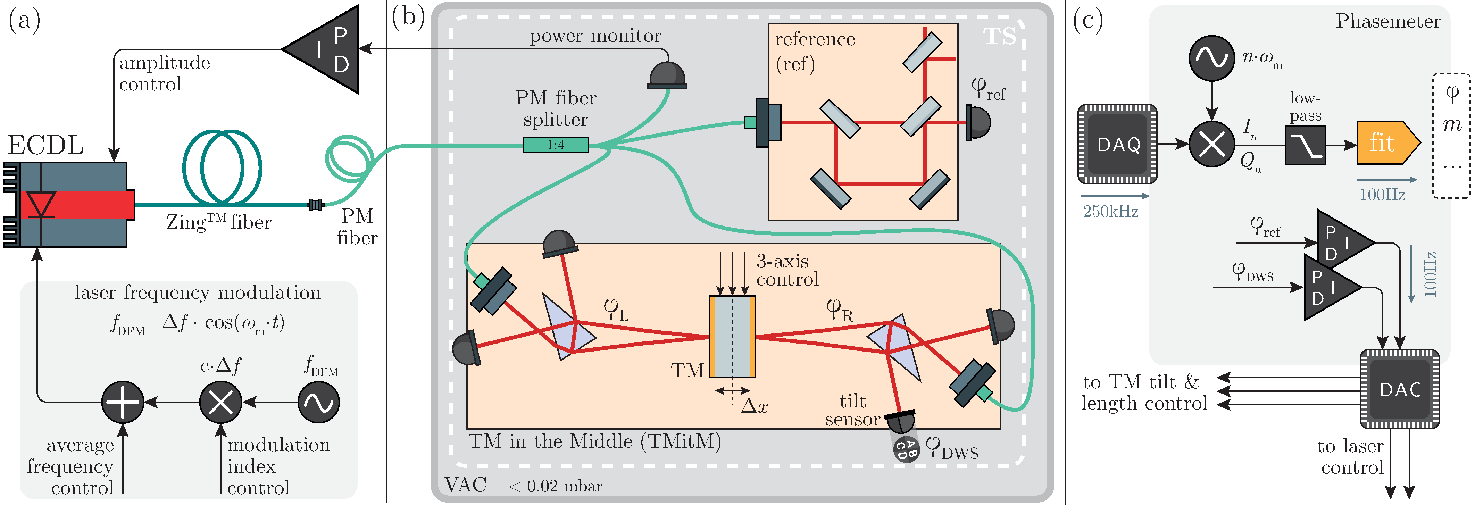}% Here is how to import EPS art
\caption{\label{fig:DFM_experimentOverview_lab}Sketch of the experiment. Inset (a) shows the laser preparation. A deep frequency modulation, $f_\mathrm{DFM}$, is applied to a fiber-coupled external cavity diode laser (ECDL). All fiber components are polarisation maintaining (PM). The vacuum chamber (VAC), shown in inset (b), houses a thermal shielding (TS) which covers two quasi-monolithic interferometers: the reference (ref) and the Test Mass in the Middle (TMitM). The data acquisition (DAQ) system has eight analogue inputs which are simultaneously digitised and processed in a software phasemeter, both shown in inset (c). Analogue control voltages are provided by a digital-to-analogue converter (DAC) and are used to actuate on the laser and test mass (TM). }
\end{figure*}

%%%%%%%%%%%%%%%%%%%%%%%%%%%%%%%%%%%%%%%%%%%%%%%%%%%%%%%%%%%%%%%%%%%%%%%%%%%%%%%%%%
\section{Experiment}
%%%%%%%%%%%%%%%%%%%%%%%%%%%%%%%%%%%%%%%%%%%%%%%%%%%%%%%%%%%%%%%%%%%%%%%%%%%%%%%%%%
DFMI uses a single laser beam that is strongly modulated sinusoidally in its frequency. By injecting this light into an interferometer with unequal arms as shown in Fig.~\ref{fig:DFM_experimentOverview_lab}, we generate an output pattern that contains complex amplitudes at the modulation frequency and its harmonics. The essence of DFMI, as we use it, is the phase extraction algorithm: We employ a non-linear fit algorithm based on a Levenberg-Marquardt (least squares) routine. It uses  a Bessel function decomposition of the complex amplitudes at the modulation frequency and its first, about ten, harmonics \cite{heinzel2010deep}. This allows us to extract four fundamental measurement parameters in real-time: the interferometric phase $\varphi$ and amplitude, the modulation index $m$ and the modulation phase as shown in Fig.~\ref{fig:DFM_experimentOverview_lab}(c). 
The effective modulation index depends on the frequency modulation (FM) deviation, $\Delta f$, and on the interferometer arm length mismatch, $\Delta L$. It is given by the relation $m = 2 \pi \Delta f \Delta L/c_0$, with the speed of light $c_0$. By increasing the FM deviation the size of the optical set-ups can be reduzed. Estimates of absolute distances are also possible with this readout scheme by tracking the modulation index $m$ \cite{Gerberding2015}.

\subsection{Laser}
As illustrated in Fig.~\ref{fig:DFM_experimentOverview_lab}(a), the laser source is a fiber-coupled external cavity diode laser (ECDL) with a center wavelength of $\lambda = 1064\,\mathrm{nm}$ that is modulated by $\Delta f \approx \pm 5\,\mathrm{GHz}$ with a rate of $f_\mathrm m = 0.8\,\mathrm{kHz}$. It provides 15\,mW optical power that is routed into a vacuum chamber (shown in inset (b)).  
A polarization maintaining fiber splitter equally distributes the light inside the chamber via four ports to a power monitor, the so-called Test Mass in the Middle (TMitM) experiment and another quasi-monolithic interferometer which is used as frequency reference. This optical reference allows us to measure the FM deviation and the laser frequency noise accumulated over 7\,cm arm length mismatch and to stabilise both \cite{gerberding2017laser}. 
Amplitude fluctuations, caused by the strong frequency modulation and polarisation fluctuations in the fibers, are actively stabilised by a closed loop control of the laser diode current with $100\,\mathrm{kHz}$ unity gain frequency.

% 188.736 mm x 68.086 mm 
% ratio = 2.7720236172
% words = 300 / (0.5 * 2.7720236172) + 40 = 256.4483723288
 \begin{figure*}[]
\centering
\includegraphics[width=0.99\textwidth]{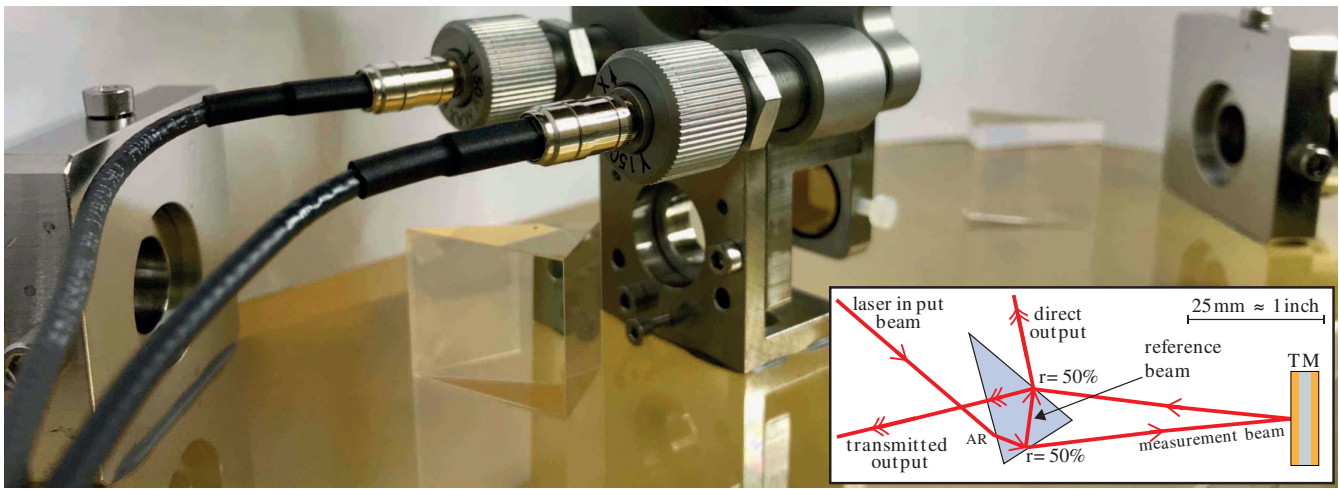} % Here is how to import EPS art
%\caption{\label{fig:img1}} 
\hfill
\caption{\label{fig:img1a}Photograph of the TMitM. One prism is placed on both sides of the gold-coated mirror (12.7\,mm in diameter), which is suspended from a three-axes piezo actuator. Titanium fiber collimators fixed in Invar holders provide light to each interferometer. The inset shows a sketch of the left prism-shaped interferometer with beam splitter and anti-reflective (AR) coating. The laser beam paths are drawn in red and hit the test mass (TM) under 4.1$^\circ$.}
\end{figure*}

%%%%%%%%%%%%%%%%%%%%%%%%%%%%%%%%%%%%%%%%%%%%%%%%%%%%%%%%%%%%%%%%%%%%%%%%%%%%%%%%%%
\subsection{Test Mass in the Middle (TMitM)}
%%%%%%%%%%%%%%%%%%%%%%%%%%%%%%%%%%%%%%%%%%%%%%%%%%%%%%%%%%%%%%%%%%%%%%%%%%%%%%%%%%
The TMitM is the key part of this experiment to test DFMI. A photograph of the interferometers is shown in Fig.~\ref{fig:img1a}. It consists of a 4\,mm thick mirror that is gold-coated on both sides and mounted on a three-axis piezo-transducer (PZT) that is glued in the center of an optical bench (OB). The OB is made of the glass-ceramic Clearceram with a low coefficient of thermal expansion of $1\cdot 10^{-8}\,/\mathrm K$. Two interferometers, one on each TM side, allow us to perform two redundant interferometric measurements of the same TM motion.

For this, a single-component, prism-shaped, interferometer was designed by using the optical simulation tool \textsc{ifocad} \cite{Wanner12OC}. It has a triangular base surface with two equal sides of about 25\,mm length. The dimensions are chosen such that commercially available prisms made of Fused Silica can be used. Only the perpendicularity between the optical surfaces and base is critical. It must be better than 2" to ensure a sufficient vertical alignment of a monolithic interferometer. The internal prism angles are compensated during alignment by prism rotation and input beam tilts, requiring no stringent production tolerances or excessive alignment procedures. Tolerable TM rotations before loss of heterodyne efficiency are also comparable to classic interferometers with the same beam parameters. In our case this corresponds to a heterodyne efficiency of more than 30\% for $\pm 250\,\upmu\mathrm{rad}$ tip and tilt angles and $\pm 2\,\mathrm{mm}$ TM displacement. 
% The first optical surface of the prism has an anti-reflective coating, the second surface has a beam splitter coating with 50\% reflectivity, such as the third surface, which is used for recombining reference and measurement beam. 
 A sketch of the component with laser propagation and coatings is shown in inset of Fig.~\ref{fig:img1a}. The reference beam travels a short distance inside the medium while the measurement beam leaves the medium and is reflected by the TM. The interferometer is designed for an angle of incidence of $4.1^\circ$, an arm-length mismatch of 83.3\,mm and it provides two redundant optical output ports which we call \textit{direct} and \textit{transmitted}. The transmitted port contains the interference signal with a phase shift of $\pi$ with respect to the direct one. The transmitted beams also leave the medium without compensating the previous refraction which leads to beam profile distortions in the transmitted port and an effective ellipticity of about 0.7 for both interfering beams.

Two prisms are glued with UV cured adhesive onto the OB of the TMitM. Two adjustable fiber collimators provide the light for the prisms. Both interferometers are aligned such that after vacuum chamber evacuation we achieve maximum contrast (more than 95\%), taking into account a small alignment change due to the refractive index of air.

 \subsection{Readout and control}
 We count three interferometers inside the vacuum chamber which must be read out. Each interferometer has a slightly different modulation index due to its individual arm length mismatch. The reference interferometer's modulation index is $m_\mathrm{REF} = 7.5\,\mathrm{rad}$ (stabilised). The two prims interferometers in the TMitM show an index of about $m_\mathrm{L}\approx 8.9\,\mathrm{rad}$ in the left and $m_\mathrm{R}\approx 9.1\,\mathrm{rad}$ in the right one.
 
 The readout algorithm of DFMI is briefly illustrated in Fig.~\ref{fig:DFM_experimentOverview_lab}(c). The extraction of the desired fit parameters is done in real-time in the \textit{phasemeter} which is implemented in a \textsc{c}-program running on a PC. An 8-channel data acquisition card with 250\,kHz sampling rate is used to digitise the photodiode voltages, which were generated from the photo currents via transimpedance amplifiers. A digital-to-analogue converter provides actuation signals from the phasemeter via a USB interface. Five digital feedback control loops (proportional-integral controllers) are integrated in software: The tilt of the TM in horizontal and vertical direction can be controlled by actuating on the three-axis PZT mount. As error signal we use a differential wavefront sensing (DWS) measurement which is provided by quadrant photodiodes (QPDs). 
 The limited number of readout channels allowed us to only use one QPD during displacement performance measurements, which we used to actively suppress TM tilts.  
  The coupling factor between optical tilt and DWS signal of about $5000\,\mathrm{rad}/\mathrm{rad}$ was calculated by means of optical simulations for an assumed beam waist radius of 0.5\,mm and a waist position of 100\,mm; lenses with 25.4\,mm focal length produced a spot size radius of 0.12\,mm on the diodes' active area. The experimental data verified this coupling factor. The pathlength of the TM is controlled by actuating on all three PZT axes simultaneously.  
The two remaining feedback loops control the average laser frequency and its applied FM deviation by tuning the modulation DC voltage and AC amplitude, respectively.  The software phasemeter restricts the bandwidths of all control loops to about 10\,Hz.

%%%%%%%%%%%%%%%%%%%%%%%%%%%%%%%%%%%%%%%%%%%%%%%%%%%%%%%%%%%%%%%%%%%%%%%%%%%%%%%%%%
\section{Results}
%%%%%%%%%%%%%%%%%%%%%%%%%%%%%%%%%%%%%%%%%%%%%%%%%%%%%%%%%%%%%%%%%%%%%%%%%%%%%%%%%%
% current measurement: 		2018_07_26_17_57_17
% dynamic range measure: 	

%\begin{figure*}[]
%\centering
%%\includegraphics[width=0.99\textwidth]{prism_photo3.eps} % Here is how to import EPS art
%%\caption{\label{fig:img1}} 
%\includegraphics[width=0.99\textwidth]{prism_photo6.eps} % Here is how to import EPS art
%%\includegraphics[height=4.25cm]{IMG_7949.eps} % Here is how to import EPS art
%%\includegraphics[width=0.45\textwidth]{img_TMitM.eps} % Here is how to import EPS art
%\caption{\label{fig:img1}Photograph of the Test Mass in the Middle (TMitM) experiment. One prism is placed on both sides of the gold-coated mirror (12.7\,mm in diameter), which is suspended from a three-axes piezo actuator. Titanium fiber collimators fixed in Invar holders provide light to each interferometer.}
%\end{figure*}

The TMitM measures two interferometric phases, $\varphi_\mathrm{L}$ and $\varphi_\mathrm{R}$, on the left and on the right side of the TM. The wavenumber $k=2\pi/\lambda$ converts the phases into displacements, $\tilde x = \varphi / k$, and we can define $\tilde x_\mathrm L$ and $\tilde x_\mathrm R$ for both sides, and equivalent $\tilde x_\mathrm{ref}$ for the reference interferometer. A TM motion, $\Delta x$, is monitored redundantly in both prism interferometers except for an opposite sign. We introduce the parameter $\varepsilon$ which adds an additional noise term that is not common mode in the two interferometers. If one also includes laser frequency noise, $\nu$, it couples into all interferometers scaled by their arm length difference, given by $m_\mathrm L$, $m_\mathrm R$, $m_\mathrm{ref}$, as extracted by the phasemeter. The displacement noise for each interferometer can be described by:
\begin{eqnarray}
	\tilde x_\mathrm L &=& +2\Delta x + \nu \cdot m_\mathrm L, \label{eq:xl}\\
	\tilde x_\mathrm R &=& -2\Delta x + 2\varepsilon + \nu \cdot m_\mathrm R, \label{eq:xr}\\
	\tilde x_\mathrm{ref} &=& \sigma + \nu\cdot m_\mathrm{ref}.\label{eq:xref}
\end{eqnarray}
The test mass displacement, $\Delta x$, occurs amplified by a factor of 2 in each prism measurement due to the reflection set-up. With an angle of incidence of about 4.1$^\circ$ this results in a coupling factor of nearly 2 which is inserted in the equations above. 
The measurement of the reference interferometer contains an additional phase noise term, $\sigma$, that is driven by e.g. wrong polarizations or other noise influences unique to this interferometer \cite{gerberding2017laser}. 
Eqs.~(\ref{eq:xl})-(\ref{eq:xref}) can be combined to two expressions:
\begin{eqnarray}
	%\tilde x_\mathrm L + \tilde x_\mathrm R &=&  2\tilde\nu + 2 \varepsilon, \\
	\tilde x_\mathrm{TM} &:= & (\tilde x_\mathrm L - \tilde x_\mathrm R)/4 \approx \Delta x + \varepsilon/2, \label{eq:xtm}\\
	\tilde x_\mathrm{RPN} &:=&		(\tilde x_\mathrm L + \tilde x_\mathrm R - \tilde x_\mathrm{ref}\cdot \rho)/4  =   \varepsilon/2 +  \sigma \cdot \rho / 4, \label{eq:rpn}
\end{eqnarray}
with $\rho = (m_\mathrm L + m_\mathrm R) / m_\mathrm{ref}$. Eq.~(\ref{eq:xtm}) contains the actual TM motion and the noise term $\varepsilon$. A minor coupling of residual frequency noise, $\mathcal O(m_\mathrm L -m_\mathrm R)$, is present but was found to be negligible here. Eq.~(\ref{eq:rpn}) combines all three measurements to reveal the total residual phase noise (RPN), containing the two noise terms $\sigma$ and $\varepsilon$. The RPN can be used to analyse any residual noise in the testbed or of the readout technique itself. 
%This equation gives an estimation on the total residual phase noise (RPN) by making the sum of the left and right prism interferometer. The phase from the reference interferometer is subtracted subsequently to eliminate residual laser frequency noise. We use the effective modulation indices to scale the laser frequency noise according to the different arm length mismatches with the factor $\rho = (m_\mathrm L + m_\mathrm R) / m_\mathrm{ref}$. 
A further signal combination is the opto-electronic noise (OEN), $\tilde x_\mathrm{OEN}$, that shows the difference of two redundant outputs of a recombination beam splitter, e.g. $\tilde x_\mathrm{L,direct}-\tilde x_\mathrm{L,transmitted}$, which is a useful quantity for the evaluation of readout or stray light noise in the experiment.

The result of the TM displacement noise is plotted in Fig.~\ref{fig:result1}. By applying all control schemes (laser frequency/modulation and TM stabilization) we are able to sense TM displacements with $230\,\mathrm{fm}/\sqrt{\mathrm{Hz}}$ precision between 300\,mHz and 10\,Hz ($\tilde x_\mathrm{RPN}$). The TM motion, $\tilde x_\mathrm{TM}$, shows artefacts at higher harmonics that are induced by acoustic and seismic noise. The peak at 25\,Hz was identified to be the oscillation of the vacuum pump running during this measurement. At frequencies below 30\,mHz the noise increases with a $1/f^2$-behaviour. The reasons for this are to some extent temperature fluctuations that cause beam jitter, or other non-linear effects.

\subsection{Non-linearities}
\begin{figure}[]
\centering
	\includegraphics[]{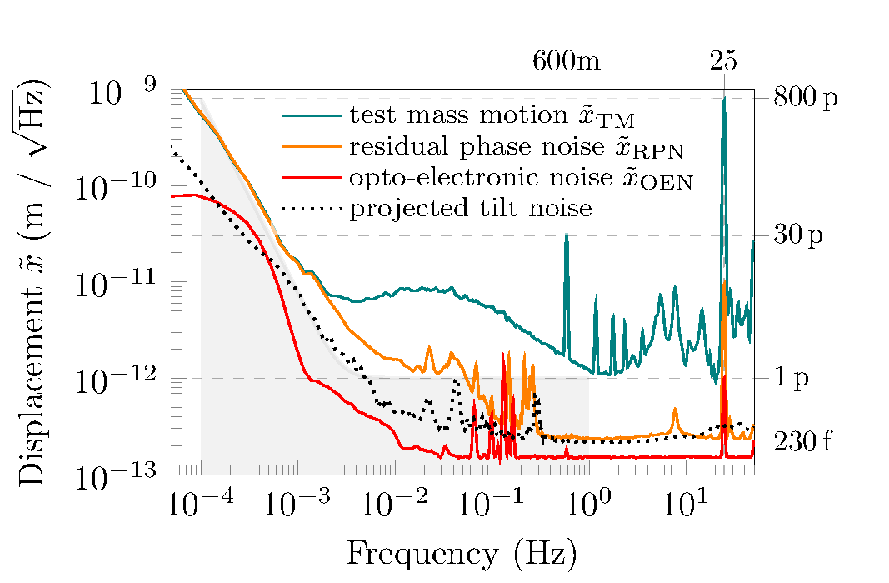} 
	\caption{\label{fig:result1}Displacement performance of DFMI. As reference we plot the sensitivity curve of the LISA mission concept document of $1\,\mathrm{pm}/\sqrt{\mathrm{Hz}}$ relaxed to frequencies below 3\,mHz as grey area \cite{LISAmissionL3}. The individual signals are calculated from the interferometric phases according to Eqs.~(\ref{eq:xtm}) and(\ref{eq:rpn}). Shot noise is not visible in this plot and expected to be below $10^{-14}\,\mathrm{m}/\sqrt{\mathrm{Hz}}$.}
\end{figure}

Amplitude fluctuations couple into the opto-electronic noise, $\tilde x_\mathrm{OEN}$, and might be caused by the laser source itself or due to polarisation fluctuations. The amplitude stabilisation reduces this noise up to a bandwidth of 100\,kHz. Digitisation noise limits $\tilde x_\mathrm{OEN}$ at $150\,\mathrm{fm}/\sqrt{\mathrm{Hz}}$ down to 30\,mHz, as shown in Fig.~\ref{fig:result1}. We assume that polarisation fluctuations caused by refractive index changes in the fiber components are the main reason for the increasing noise level below 30\,mHz. A residual amplitude modulation caused by the applied strong frequency modulation, and not sufficiently suppressed by the stabilisation, cannot be excluded either. 

The reference interferometer is used as sensor to stabilise the laser frequency noise and to lock the modulation index to a constant value, here $m_\mathrm{ref} = 7.5\,\mathrm{rad}$. The stabilisation of the modulation index is useful to lock the operation point of the laser. While the control of the modulation index did not show any significant improvements in the phase performance, we are able to reduce laser frequency noise drifts below the unity gain frequency of 10\,Hz. By subtracting the remaining laser frequency noise in data post-processing, we are able to further improve the performance between 100\,mHz and 50\,Hz by one order of magnitude at 1\,Hz. The resulting residual phase noise is given by $\tilde x_\mathrm{RPN}$ in Fig.~\ref{fig:result1}. It is limited by OEN above 1\,Hz and by some unexplained non-common mode noise source below this frequency. Despite this excess noise the measurement shows that DFMI with single-component interferometers is able to achieve 1\,pm-level precision comparable to, and partly better than, previous breadboard experiments for LISA and LPF \cite{heinzel2004ltp,de2010experimental,schuldt2012picometre,watchi2018contributed}.

% 88.171 mm x 51.059 mm 
% ratio = 1.7268454141
% words = 150 / 1.7268454141 + 20 = 106.8635946074

The measurement also shows a dynamic range of more than two orders of magnitude at 600\,mHz and 25\,Hz between $\tilde x_\mathrm{TM}$ and $\tilde x_\mathrm{RPN}$. 
This is compliant with gravity missions that use some form of feedback control such as LISA, LISA Pathfinder or GOCE-type instruments.
Residual readout non-linearities might be caused by the strong frequency modulation itself that could limit the dynamic range capabilities, as discussed in previous studies \cite{Isleif:16,Gerberding2015}. Possible reasons are harmonic distortions of the modulation caused by the function generator or the laser itself via the external cavity excitation. In the current range these effects do not seem to be limiting, however, they might become dominant for higher dynamics if no further compensation is applied. The influence of non-flat transfer functions of the photo-receivers was also measured and corrected and is most probably not the limiting factor at this point. Other geometrical effects, like non-parallel incidence or a wedge of the TM, can also lead to non-linear couplings. The current noise limitations are however more likely explained by undesired tilts as described in the following.

\subsection{Tilt-to-length noise}

% 85.885 mm x 47.414 mm 
% ratio = 1.811384823
% words = 150 / 1.811384823 + 20 = 102.8095709401
\begin{figure}[]
\centering
	\includegraphics[]{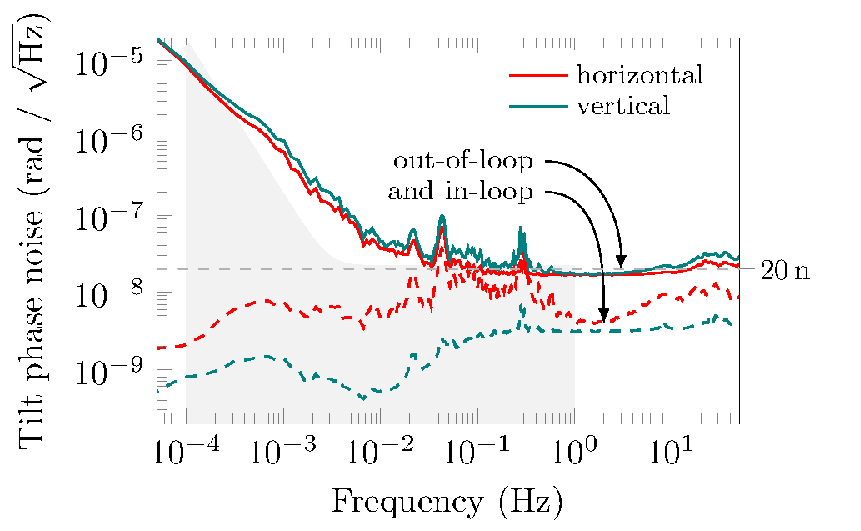} 
	\caption{\label{fig:result2}Tilt noise performance of DFMI. As reference we plot the mission requirements of the LPF test campaign of $20\,\mathrm{nrad}/\sqrt{\mathrm{Hz}}$ relaxed to frequencies below 3\,mHz as grey area \cite{cervantes2013lisa}. To get the optical tilt noise performance we divided the measured DWS phase measurement by the coupling factor ($\approx$5000\,rad/rad).}
\end{figure}

In a subsequent measurement we used a second QPD in the other prism interferometer and monitored the out-of-loop behaviour of TM tilts. Due to the limited number of readout channels we were not able to stabilise the laser frequency and modulation nor the TM pathlength during this additional measurement. Only the TM tilt noise was monitored and the according in- and out-of-loop measurements are shown in Fig.~\ref{fig:result2}. 
We are able to stabilise the TM tilt down to noise levels on the order of $20\,\mathrm{nrad}/\sqrt{\mathrm{Hz}}$ above 40\,mHz, as the out-of-loop measurements indicate. The projected TM displacement noise can be determined by combining the in-loop and out-of-loop tilt measurements. The DWS and tilt-to-length coupling factors for beam jitter were determined by optical simulations and are about 5000\,rad/rad and 4\,pm/$\upmu$rad, including some initial interferometric misalignment of about $50\,\upmu\mathrm{rad}$. The projected tilt noise is shown by the dotted black line plotted in Fig.~\ref{fig:result1}. It matches the residual phase noise, $\tilde x_\mathrm{RPN}$, above 200\,mHz and might explain some of the noise increase below 200\,mHz. Not measured here are beam pointing fluctuations of the reference interferometer which likely also drive the performance of $\tilde x_\mathrm{RPN}$ by coupling into the measurement via residual noise of the laser frequency stabilisation \cite{gerberding2017laser}. This could explain the missing correlation between projected tilt noise and residual phase noise measurement at low frequencies. 

%%%%%%%%%%%%%%%%%%%%%%%%%%%%%%%%%%%%%%%%%%%%%%%%%%%%%%%%%%%%%%%%%%%%%%%%%%%%%%%%%%
\section{Summary}
%%%%%%%%%%%%%%%%%%%%%%%%%%%%%%%%%%%%%%%%%%%%%%%%%%%%%%%%%%%%%%%%%%%%%%%%%%%%%%%%%%
The usage of deep frequency modulation interferometry (DFMI) allows simpler interferometer set-ups in comparison to heterodyne interferometry. Especially it scales more easily to multiple TM displacement and tilt readouts on the same or adjacent ultra-stable benches. It can further maintain multi-fringe capabilities that are required, for example, for a TM readout in space. Only one frequency reference interferometer is required per laser source. The so-prepared laser light can be distributed into many optical set-ups whose number is finally constrained by the minimal optical power required not to be limited by shot noise.

Using a compact, single-component beam splitting and recombination optic specifically designed for DFMI we conducted a prototype test mass experiment in which we were able to reach displacement sensing levels of $230\,\mathrm{fm}/\sqrt{\mathrm{Hz}}$ around 300\,mHz and tilt noise readouts with sensitivities of about $20\,\mathrm{nrad}/\sqrt{\mathrm{Hz}}$ at 40\,mHz. The current performance limitation at low frequencies is most probably caused by beam pointing fluctuations which can be reduced by using ultra-stable, monolithic fiber collimators in future experiments. At high frequencies we are currently limited by the digitisation noise of our data acquisition system. The required strong frequency modulation causes non-linearities which were discussed and found not to be limiting in the dynamic range regime probed here. Future work will concentrate on the development of scalable phasemeters with many channels and on optimising the readout and laser control algorithms to further test and increase the phase readout linearity. The experiment represented here is a well-suited testbed for these future investigations.

To conclude, the simplicity of the compact interferometric setup and the achieved LISA-like performance make deep frequency modulation interferometry attractive for future experiments and metrology applications.

%We have shown the design of a compact single-component interferometer and experimentally investigated a prototype test mass experiment. 
%The strong frequency modulation causes certain non-linearities which were discussed in this article and found not to be limiting in the here probed dynamic range regime. We experimentally validated that deep frequency modulation interferometry is able to reach  LISA-like displacement performances of $230\,\mathrm{fm}/\sqrt{\mathrm{Hz}}$ around 300\,mHz and  tilt noise readouts with sensitivities of about $20\,\mathrm{nrad}/\sqrt{\mathrm{Hz}}$ at 40\,mHz. The simplicity of the compact interferometric set-up and the achieved performances makes this technique also attractive for other experiments and metrology applications.

% The experiment represented here is a well-suited testbed for these future investigations.

\begin{acknowledgments}
The authors would like to thank the DFG Sonderforschungsbereich (SFB) 1128 Relativistic Geodesy and Gravimetry with Quantum Sensors (geo-Q) and the International Max-Planck Research School (IMPRS) for financial support.
\end{acknowledgments}

%\bibliography{../Bibliography}


%\begin{thebibliography}{27}%
%\makeatletter


%\end{thebibliography}%

\end{document}